\documentclass[11pt,a4paper]{article}
\pdfoutput=1
\usepackage{jcappub}

\usepackage{color}
\input{colordvi.tex}
\usepackage{amsmath}
\usepackage{graphicx}
\usepackage{float}
\usepackage{wrapfig}
\usepackage{bm}
\usepackage{amsmath}
\usepackage{ltgtsim}
\usepackage{braket}

\usepackage{amsmath,amssymb}
\usepackage{graphicx}
\usepackage{subfigure}
\usepackage{verbatim}
\usepackage{makeidx}
\usepackage{amssymb}
\usepackage{ulem}

\title{Anisotropic CMB distortions from non-Gaussian isocurvature perturbations}

\author[a]{Atsuhisa Ota,}
\author[b]{Toyokazu Sekiguchi,}
\author[c,d]{Yuichiro Tada}
\author[e]{and Shuichiro Yokoyama}
\affiliation[a]{Department of Physics, Tokyo Institute of Technology,\\
Tokyo 152-8551, Japan}
\affiliation[b]{University of Helsinki and Helsinki Institute of Physics,\\ P.O. Box 64, Helsinki 00014, Finland}
\affiliation[c]{Kavli Institute for the Physics and Mathematics of the Universe (WPI), The University of Tokyo,\\ Kashiwa, Chiba 277-8583, Japan}
\affiliation[d]{Department of Physics, the University of Tokyo,\\ Bunkyo-ku 113-0033, Japan}

\affiliation[e]{Department of Physics, Rikkyo University,\\ 3-34-1 Nishi-Ikebukuro, Toshima, Tokyo
223-8521, Japan}

\emailAdd{a.ota@th.phys.titech.ac.jp}
\emailAdd{toyokazu.sekiguchi@helsinki.fi}
\emailAdd{yuichiro.tada@ipmu.jp}
\emailAdd{shuichiro@rikkyo.ac.jp}

\abstract{
We calculate the CMB $\mu$-distortion and the angular power spectrum of its cross-correlation with the temperature anisotropy
in the presence of the non-Gaussian neutrino isocurvature density (NID) mode.
While the pure Gaussian NID perturbations give merely subdominant 
contribution to $\langle\mu\rangle$ and vanishing $\langle \mu T\rangle$,
the latter quantity can be large enough to be detected in the future
when the NID perturbations $\mathcal S(\bm x)$ are proportional to the square of a Gaussian field $g(\bm x)$, i.e. $\mathcal S({\bm x})\propto g^2({\bm x})$.
In particular, large $\langle \mu T\rangle$ can be realized since Gaussian-squared perturbations can yield
a relatively large bispectrum, satisfying the constraints from the power spectrum of CMB anisotropies, i.e. $\mathcal{P}_\mathcal{SS}(k_0) \sim\mathcal{P}_g^2(k_0)\ltsim10^{-10}$ at $k_0=0.05$~Mpc$^{-1}$.
We also forecast constraints from the CMB temperature and E-mode polarisation bispectra, 
and show that $\mathcal{P}_g(k_0)\ltsim10^{-5}$ would be allowed from Planck data.
We find that $\langle \mu \rangle$ and $|l(l+1)C^{\mu T}_l|$ can respectively
be as large as $10^{-9}$ and $10^{-14}$ with uncorrelated scale-invariant NID perturbations
for $\mathcal{P}_g(k_0)=10^{-5}$.
When the spectrum of the Gaussian field is blue-tilted (with spectral index $n_g \simeq 1.5$), 
$\langle \mu T\rangle$ can be enhanced by an order of magnitude.
}

\begin{document}

\begin{flushright}
IPMU 14-0357\\
RUP-14-20
\end{flushright}

\maketitle
\flushbottom

\section{Introduction}

Inflationary scenario is a successful prescription to solve the initial condition problems of the hot Big Bang universe, and to determine the concrete theoretical model is one of the most important theme of recent observational cosmology~\cite{inflation} .
Such an accelerated expanding universe predicts generation of the almost scale-invariant and Gaussian primordial fluctuations, which are well confirmed by a variety of cosmological observations
including anisotropies in the Cosmic Microwave Background (CMB).
The next step for further refinement may be
to investigate deviations from Gaussian statistics, that is, to find non-Gaussianity~\cite{Maldacena:2002vr}.
The newest space mission Planck revealed that the non-linearity parameter 
$f_\mathrm{NL}\ltsim\mathcal{O}(1)$, and many inflationary models are ruled out~\cite{Ade:2013uln}.
However, we should note that the CMB
anisotropies are observables on relatively larger scale up to the multipole $l \sim 10^{4}$, 
and several orders of improvement of the resolution can be far beyond our current technology as well.

CMB spectrum distortions are alternative probes of
the inflationary universe, for the distortions are created from small-scale density perturbations, without introducing new physics.
Use of CMB distortions in conjunction with the CMB anisotropies on large scales may complementarily allow us to 
test the nature of primordial perturbations over many orders of scales.
The distortions are basically classified into two types, $y$-type and $\mu$-type,\footnote{
Some people discuss the intermediate type distortions~\cite{Burigana:1995,Khatri:2012tw}.}
depending on whether the system is thermal or not~\cite{Zeldovich:1969ff,1970Ap&SS...7...20S}.
Although thermalization of the photon system is efficient in the early universe, once the Compton scattering becomes ineffective around 
the redshift $z\sim 10^5$, deviations from the
thermal equilibrium can no longer vanish
since the Thomson scattering never transfers the photon energy through non-relativistic electrons.
Then the distortion is parameterized as a non-thermal deviation from Planck distribution and we call it $y$-distortion.
Another one is $\mu$-distortion, which is a chemical potential in the
Bose-Einstein distribution function of thermal photon system.
Such a thermal deviation from Planck distribution is a consequence of re-thermalization under number conserving process such as the Compton scattering, so we can investigate the Compton dominant era around $z \sim 10^6$~\cite{Danese:1982,1991A&A...246...49B,Hu:1994bz,Chluba:2006kg,Khatri:2012tv,Burigana:1995,Khatri:2012tw}.
Here we would focus only on this type of distortions.
Typical magnitude of the $\mu$-distortion which originates from the primordial curvature perturbations
is $10^{-8}$~\cite{Hu:1994bz,Sunyaev:1970eu,
1991MNRAS.248...52B,1991ApJ...371...14D} and the contributions from primordial tensor perturbations are subdominant~\cite{Ota:2014hha,Chluba:2014qia}.
Constraints on these parameters are given by COBE FIRAS as $\mu<9\times 10^{-5}$
and $y<1.5\times 10^{-5}$ (95\% C.L.)~\cite{Mather:1993ij,Fixsen:1996nj,Salvaterra:2002mg}, and future space mission such as PIXIE~\cite{Kogut:2011xw} and
PRISM~\cite{Andre:2013afa} have potential to improve the constraints up to the order of $10^{-8}$ to $10^{-9}$.
Therefore, such distortions can be
powerful tools to study primordial fluctuations.

Recently $\mu T$ cross-correlation is also proposed as a probe of
primordial non-Gaussianities down to small-scales~\cite{Pajer:2012vz,Ganc:2012ae}.
Roughly speaking, $\mu$ is proportional to the square of dimensionless temperature perturbations and then, the cross-correlation originates from the primordial
three-point function.
For the local type non-Gaussianity,
the constraints on the non-linear parameter by PIXIE's sensitivity are estimated as $f^{\rm loc}_{\rm NL}\lesssim 10^3$ in Ref.~\cite{Pajer:2012vz}.
In this paper we calculate the $\mu T$ cross-correlation in the presence of not only adiabatic but also isocurvature modes.
The isocurvature modes do not necessarily follow the Gaussian distribution unlike the adiabatic modes
and then they can cause large $\mu T$ signals. Accordingly,
we calculate contribution from Gaussian-squared isocurvature perturbations to the total cross-correlation
as an example of the non-Gaussian case.  
Here we focus on the neutrino isocurvature density (NID) mode, which is not suppressed on small scales compared to adiabatic perturbations, 
in contrast to the matter isocurvature perturbations~\cite{Dent:2012ne,Chluba:2013dna}. 
We organize this paper as follows.
In section~\ref{power spectrum and bispectrum}, 
we present the formalisms of the power spectrum and bispectrum of the non-Gaussian 
isocurvature perturbations, with particular focus on the Gaussian-squared ones.
Sections~\ref{CMB mu distortion} and \ref{muT angular cross correlation} show the calculations for $\mu$ and $\mu T$ cross-correlations, respectively.
We summarize the discussions and conclude in the final section.
In appendix~\ref{app2}, we comment on the constraints from CMB angular bispectrum.

\section{Gaussian-squared type isocurvature perturbations
%Gaussian-squared type power spectrum and bispectrum
}
\label{power spectrum and bispectrum}

Models of generating the NID mode have been proposed in the 
literature~\cite{Lyth:2002my,Kawasaki:2011rc,Kobayashi:2011hp}.
So far, most of the models can be categorised into two types.\footnote{
In general, there should also be induced isocurvature perturbations between matter and radiation, 
depending on the details of production mechanisms (e.g. thermal or non-thermal ones) of baryon and CDM.
However, as we mentioned in Introduction, when we focus on signatures in the CMB $\mu$-distortion from isocurvature perturbations, 
contribution from matter isocurvature perturbations should be smaller than that from NID ones~\cite{Dent:2012ne,Chluba:2013dna}.
Therefore for simplicity we omit the matter isocurvature perturbations throughout this paper, although
there is a possibility that presence of them can somewhat change the bounds on 
non-Gaussian curvature and NID perturbations from the CMB bispectrum presented in Appendix~\ref{app2}.}
In one type, there assumed to be a large lepton asymmetry ($n_L/n_\gamma=\mathcal O(0.01)\gg n_B/n_\gamma=\mathcal O(10^{-9})$) in the universe.
If fields sourcing the lepton asymmetry spatially fluctuate differently from inflaton (or in general fields reheating the universe), 
isocurvature perturbations inevitably arise between neutrino and the photon.
In the other type, there assumed to exist dark radiation in the universe other than neutrino.
In the context of structure formation, there is no distinction between dark radiation and neutrino,
and they in effect consist a single fluid of neutrino species. Therefore, the NID mode is sourced 
if dark radiation is produced from fields which have isocurvature perturbations.

The NID mode can be non-Gaussian when so are source fields in themselves.
In addition, even when the source fields are Gaussian, the so-called local-type
non-Gaussianity in the NID mode can be induced in the similar fashion as in 
e.g. the curvaton and the modulated reheating models. 
In Ref.~\cite{Kawakami:2012ke}, several concrete models generating 
the local-type non-Gaussian NID mode are discussed.
A model of the type with dark radiation can be realised by generalising the curvaton model.
When a curvaton field which creates non-Gaussian curvature perturbations
decays into dark radiation with some branching ratio, 
non-Gaussian NID perturbations should also be created.
On the other hand, a model of the type with large lepton asymmetry can be realised
in the Affleck-Dine baryogenesis~\cite{Affleck:1984fy} with Q-ball formation~\cite{Kawasaki:2002hq}.
In this case, the non-linear dependence of the amount of the lepton asymmetry on
the initial value of the Affleck-Dine field leads to non-Gaussianity in the NID mode.
Generation of non-Gaussian NID perturbations in the modulated 
reheating scenario can also be realized~\cite{Kobayashi:2011hp}.
In this paper, we in particular focus on non-Gaussian isocurvature perturbations which are proportional to 
the square of Gaussian field (See \eqref{eq:defR}-\eqref{eq:defS}).\footnote{
While we can also consider weakly non-Gaussian (i.e. local-type) NID perturbations, 
difference in results (i.e. $\mu$ and $\mu T$ cross-correlation) are rather trivial
due to the similarity in the transfer functions.
Indeed, as we will discuss at the ends of  Section~\ref{Homogeneous distortions} and \ref{muT angular cross correlation}, 
both $\mu$-distortion and $\mu T$ cross-correlation change 
only by a constant multiplicative factor from the adiabatic case~\cite{Pajer:2012vz}. }
Such the Gaussian-squared perturbations can be realized in the {\it ungaussiton} 
model~\cite{Linde:1996gt,Boubekeur:2005fj,Suyama:2008nt}.

\begin{figure}
	\begin{center}
		\includegraphics[width=8cm]{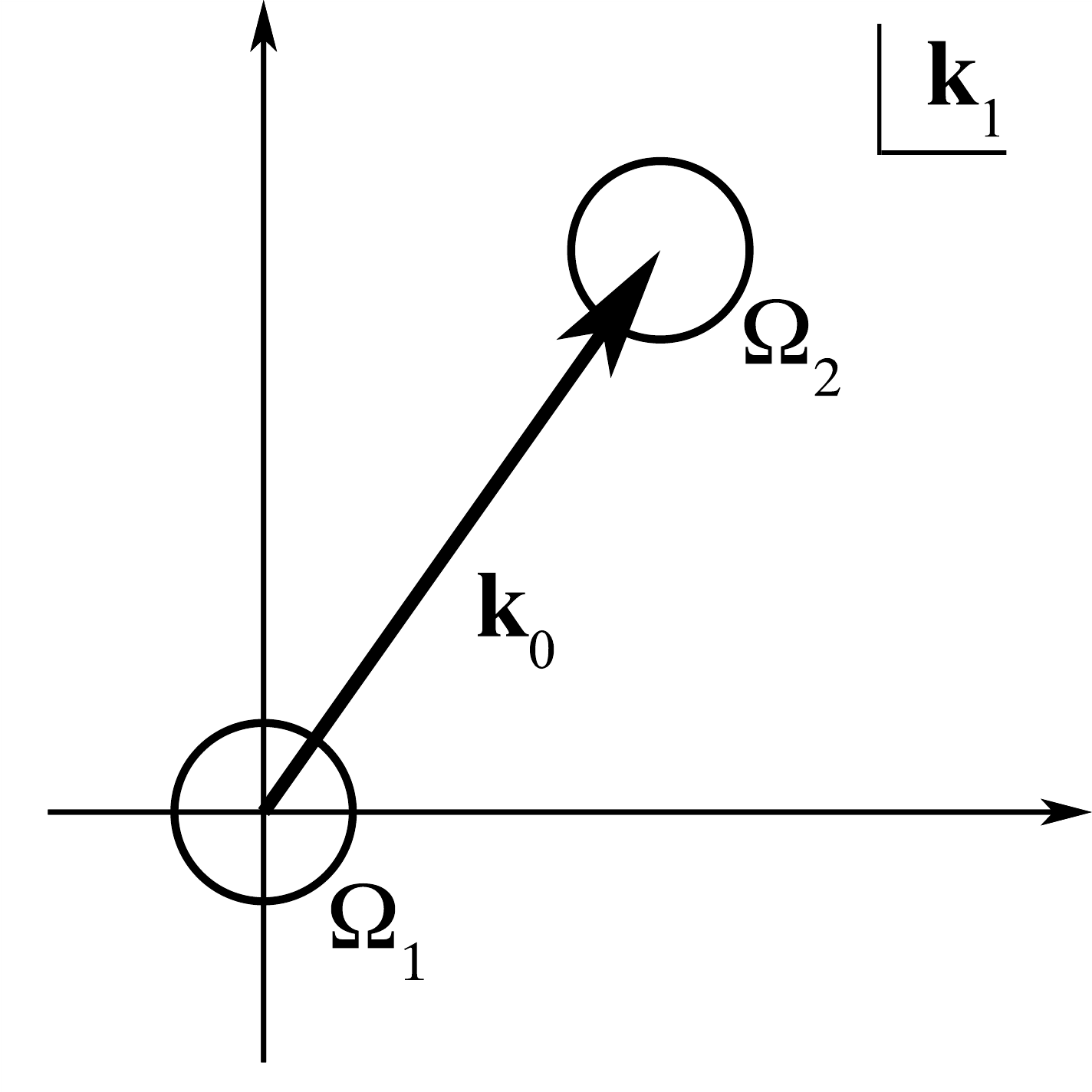}
		\caption{The schematic diagram of IR regions for the convolutional integration of $\mathcal{P}_\mathcal{SS}(k_0)$. 
		There are two IR singularities at $\bm{k}_1=0$ and $\bm{k}_0$, and the IR regions around them, $\Omega_1$ and $\Omega_2$, 
		are equivalent as they can be interchanged by variable transformation.
		It can be seen the maximum radii of them, $k_\mathrm{max}$, are about $k_0/2$.
		Hereafter we take $k_\mathrm{max}=k_0$ for simplicity.}
		\label{divergence of power}
	\end{center}
\end{figure}

\begin{figure}
\centering
\includegraphics[width=12cm]{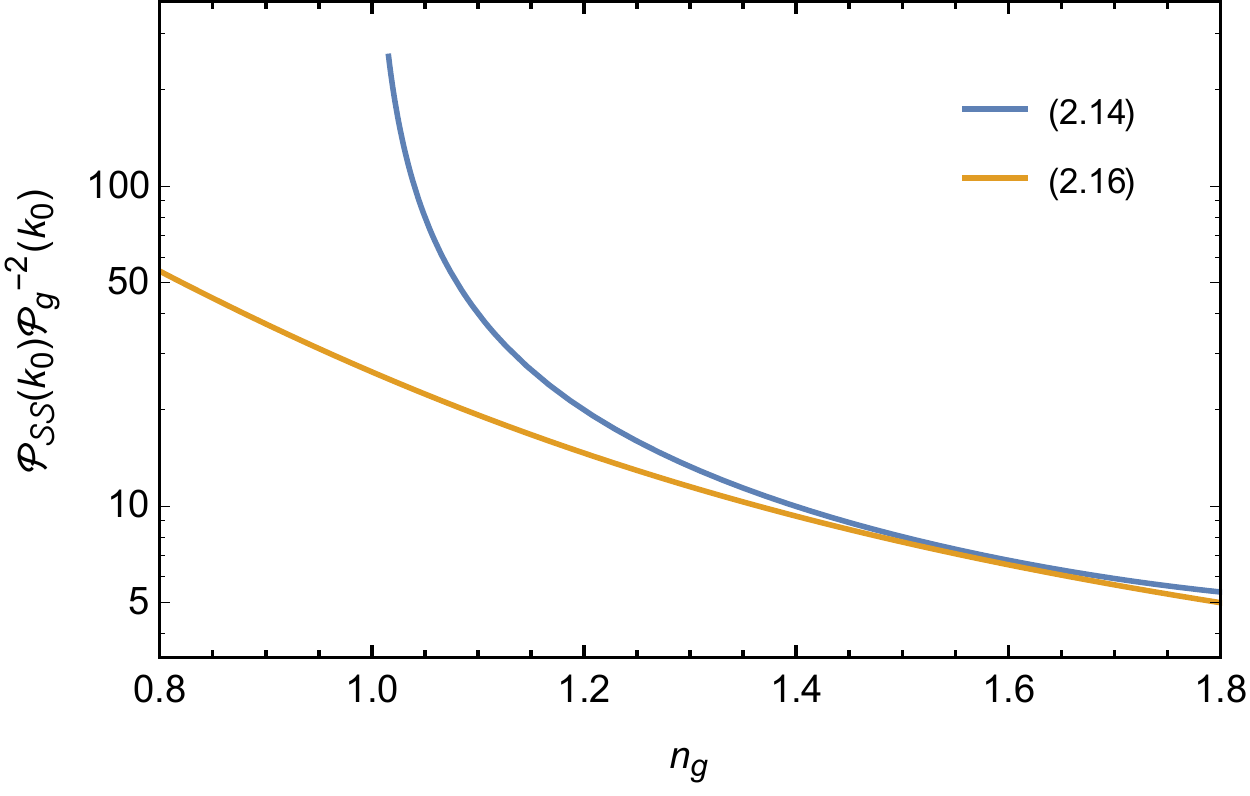}
\caption{$g^2$ isocurvature power spectrum at pivot scale $k_0 = 0.05$Mpc$^{-1}$ in units of $\mathcal P^2_g(k_0)$ with $\gamma_2=1$.
In the latter sections, we will assume $n_g\ltsim1.5$ in which the IR formulae can be used safely.}
\label{g2_iso_pow}
\end{figure}

Let us consider two 
fields $\mathcal R_G$ and $g$ both of which obey Gaussian statistics.
In the above model, 
the curvature $\mathcal R$ and the residual isocurvature perturbation 
$\mathcal S$ can be written as 
\begin{align}
\mathcal R(\bm x) &= \mathcal R_G(\bm x) + \gamma_{1} (g^2(\bm x)-\langle g^2\rangle ),\label{eq:defR}\\
\mathcal S(\bm x ) &= \gamma_{2} (g^2(\bm x)-\langle g^2\rangle ),\label{eq:defS}
\end{align}
where $\gamma_1$ and $\gamma_2$ are the model dependent constant parameters.
In this paper, we adopt the convention in Ref.~\cite{Bucher:1999re}, 
where $\mathcal S$ is defined to be the density contrast of neutrino 
in the synchronous gauge of CDM, with curvature perturbations being set to vanish.
The two-point correlation and the cross-correlation functions are 
\begin{align}
\langle \mathcal R(\bm x)\mathcal R(0) \rangle&=\langle \mathcal R_{G}(\bm x)\mathcal R_{G}(0)\rangle + 2\gamma_{1}^{2}\langle g(\bm x)g(0)\rangle^{2}, \\
\langle \mathcal R(\bm x)\mathcal S(0) \rangle&=2\gamma_{1}\gamma_{2} \langle g(\bm x)g(0)\rangle^{2},\\
\langle \mathcal S(\bm x)\mathcal S(0) \rangle &=2\gamma_{2}^{2}\langle g(\bm x)g(0)\rangle^{2},
\end{align}
and their Fourier transformations have the following form:
\begin{align}
P_{\mathcal S \mathcal S}(k)&=2\gamma_{2}^{2}\int\frac{d^{3}k_{1}}{(2\pi)^{3}}
P_{g}(k_{1})P_{g}(|\bm k- \bm k_{1}|),\label{pow:SS}\\
P_{\mathcal R \mathcal R}(k)&=P_{G}(k)+\frac{\gamma_{1}^{2}}{\gamma^{2}_{2}}P_{\mathcal S \mathcal S}(k),\\
P_{\mathcal R \mathcal S}(k)&=\frac{\gamma_{1}}{\gamma_{2}}P_{\mathcal S \mathcal S}(k),
\end{align}
where we define the power spectra $P_G$ and $P_g$ for the above two Gaussian fields,   
\begin{align}
P_{G}(k)&=\int \frac{d^{3}k}{(2\pi )^{3}}e^{-i\bm k \cdot \bm x}\langle \mathcal R_{G}(\bm x)\mathcal R_{G}(0) \rangle, \\
P_{g}(k)&=\int \frac{d^{3}k}{(2\pi )^{3}}e^{-i\bm k \cdot \bm x}\langle g(\bm x)g(0) \rangle.
\end{align}
Dimensionless power spectra are also defined as usual, i.e. $\mathcal P(k)=\frac{k^3}{2\pi^2} P(k)$
for each perturbations.
Since $\mathcal P_G$ is a power spectrum of the Gaussian curvature perturbation, we can parameterize it as
\begin{align}
\mathcal P_G(k)=A_G\left(\frac{k}{k_0}\right)^{n_s-1},
\end{align}
where $A_G=2.196\times 10^{-9}$ and $n_s=0.96$~\cite{Ade:2013uln}, with $k_0=0.05\,{\rm Mpc}^{-1}$ being the pivot scale.
The fraction
of the isocurvature perturbations is given by
\begin{align}
\beta_{\rm iso}:=
\frac{\mathcal P_{\mathcal S \mathcal S}(k_0)}{\mathcal P_{\mathcal R \mathcal R}(k_0)+\mathcal P_{\mathcal S \mathcal S}(k_0)}.
\end{align}
In the case with the neutrino density isocurvature mode,
we already have the constraints $\beta_{\mathrm{NID}}<0.27$ or 
$\mathcal P_{\mathcal{SS}}/\mathcal P_{\mathcal{RR}}<0.37$~\cite{Ade:2013uln}.
An ensemble average of a product of quantities defined as $\hat F({\bm k})=F_{\bm k}\left(\mathcal R_{\bm k} + f_{\bm k} \mathcal S_{\bm k}\right)$ is calculated as
\begin{align}
\langle \hat F(\bm k)\hat F(\bm k')\rangle=(2\pi)^{3}\delta^{(3)}(\bm k +\bm k' )F_{\bm k}F_{-\bm k}\left[
P_G(k)+\left(
f_{\bm k}f_{-\bm k}+
\frac{\gamma_1}{\gamma_2}\left(f_{\bm k}+f_{-\bm k}\right)+
\frac{\gamma^2_1}{\gamma^2_2}
\right)P_{\mathcal S \mathcal S}(k)
\right],\label{FFtransfer}
\end{align}
where $F_{\bm k}$ and $F_{\bm k}f_{\bm k}$ are transfer functions from the curvature perturbation and the isocurvature perturbation.
Here we could integrate (\ref{pow:SS}) explicitly by the use of the Feynman parameters and obtain 
\begin{eqnarray}\label{calPS:explicit}
	\mathcal{P}_{\mathcal{SS}}(k)=\gamma_2^2\mathcal{P}_g^2(k)\frac{2^{1-n_g}\pi\Gamma\left(\frac{5}{2}-n_g\right)\Gamma\left(\frac{n_g-1}{2}\right)}
	{\Gamma^2\left(2-\frac{n_g}{2}\right)\Gamma\left(\frac{n_g}{2}\right)}, \quad \mbox{for}~1<n_g<2.5,
\end{eqnarray}
where we assume
that $\mathcal P_g$ is a power of wavenumber with a spectral index $n_g$, i.e. $\mathcal P_g(k)=\mathcal P_g(k_0)(k/k_0)^{n_g-1}$.
We can see that $\mathcal P_{\mathcal S\mathcal S}(k)\propto \mathcal P_g^2(k)$, so that
$\mathcal P_{\mathcal S\mathcal S}(k)$ can be parametrized as
\begin{align}
\mathcal P_{\mathcal S\mathcal S}(k)=\mathcal P_{\mathcal S\mathcal S}(k_0)\left(\frac{k}{k_0}\right)^{2n_g-2}.
\end{align}
This suggests that the blue tilted original Gaussian field induces bluer isocurvature.
$\mathcal P_{\mathcal S\mathcal S}(k_0)$ apparently has the IR logarithmic divergence at $n_g=1$, therefore, let us introduce an IR cut-off which is motivated by the horizon size $L=14\,$Gpc to obtain a physically reasonable value for the power spectrum.
For the almost flat spectrum, most contributions are from the two IR regions, 
and they are equivalent by the translation and variable transformation (see Fig.~\ref{divergence of power}).
Then we can combine these regions into one and obtain the following form from (\ref{pow:SS}) and obtain
\begin{align}
\mathcal P_{\mathcal S \mathcal S}(k_0)&\simeq 4\gamma^{2}_{2}\mathcal P^2_g(k_0)\int^{k_{\rm max}/k_0}_{1/(k_0L)}d(\ln t)t^{n_g-1}\notag \\
&=
\frac{4}{n_g-1}\gamma^{2}_{2}\mathcal P^2_g(k_0)
\left[\left(\frac{k}{k_0}\right)^{n_g-1}\right]^{k_{\rm max}}_{L^{-1}},
\end{align}
where $k_{\rm max}$ is the upper limit of the IR region.
Expanding by $n_g-1$, we obtain the expression around the flat spectrum
\begin{align}
\mathcal P_{\mathcal S\mathcal S}(k_0) &\sim 4 \gamma^{2}_{2}\mathcal P^2_g(k_0)\log(k_{\rm{max}}L)\left[1+\frac{n_g-1}{2}\log\left(\frac{k_{\rm max}}{k^2_0 L}\right)+\cdots\right].\label{pow:ss:cutoff}
\end{align}
The IR singular points which are originally at ${\bm k}_1=0,~{\bm k}_0$ are separated by the distance of $k_0$. 
Therefore the radii of the IR regions can be taken up to about $k_0/2$ 
at most. For simplicity, we take $k_\mathrm{max}=k_0$ hereafter.
Fig.~\ref{g2_iso_pow} shows the comparison of the exact formula with the one with a IR cut-off. 
We can see that they are in good agreement up to $n_g\simeq 1.7$.

Three-point correlation functions are defined in the same manner and their bispectra can be obtained by the Fourier transformation.
For instance, the bispectrum of
the isocurvature perturbations
are written as follows.
\begin{align}
B_{\mathcal S\mathcal S\mathcal S}(k_1,k_2,k_3)&=\int d^3 x \int d^3 y e^{- i \bm k\cdot \bm x}e^{- i \bm k\cdot \bm y}\langle \mathcal S(\bm x)\mathcal S(\bm y)\mathcal S(0) \rangle\notag \\ 
&=8\gamma^3_2\int \frac{d^3 q}{(2\pi)^3}P_g(q)P_g(|\bm q - \bm k_1|)P_g(\bm q- \bm k_1 -\bm k_2|).\label{bis:SSS:int}
\end{align}
By the use of the $B_{\mathcal S\mathcal S\mathcal S}$, the other bispectra are given by
\begin{align}
B_{\mathcal R\mathcal R\mathcal R }(k_1,k_2,k_3)&=\left(\frac{\gamma_1}{\gamma_2}\right)^3B_{\mathcal S \mathcal S\mathcal S}(k_1,k_2,k_3), \\
B_{\mathcal R \mathcal R \mathcal S}(k_1,k_2,k_3)&= \left(\frac{\gamma_1}{\gamma_2}\right)^2B_{\mathcal S \mathcal S\mathcal S}(k_1,k_2,k_3),\\
B_{\mathcal R \mathcal S \mathcal S}(k_1,k_2,k_3)&= \frac{\gamma_1}{\gamma_2}B_{\mathcal S\mathcal S\mathcal S}(k_1,k_2,k_3).
\end{align}
Then a triple product of $\hat F(\bm k)$
also has the following form:
\begin{align}
&\langle \hat F(\bm k_1)\hat F(\bm k_2)\hat F(\bm k_3)\rangle \notag \\
&\quad=(2\pi )^3\delta ^3(\bm k_1+\bm k_2+ \bm k_3)B_{\mathcal S\mathcal S\mathcal S}(k_1,k_2,k_3)F_{\bm k_1}F_{\bm k_2}F_{\bm k_3}\notag \\
&\quad\quad\times \left(f_{\bm k_1}f_{\bm k_2}f_{\bm k_3} +\frac{\gamma_1}{\gamma_2}\left(f_{\bm k_1}f_{\bm k_2}+f_{\bm k_2}f_{\bm k_3}+f_{\bm k_3}f_{\bm k_1}    \right) +\frac{\gamma^2_1}{\gamma^2_2}\left(f_{\bm k_1} +f_{\bm k_2} +f_{\bm k_3}\right)  +\frac{\gamma^3_1}{\gamma^3_2}\right).
\end{align}
$B_{\mathcal S \mathcal S \mathcal S}$ also has IR singularities, and here we treat them in the same way as the power spectrum.
Let us assume a 
squeezed configuration, which we will consider in the latter sections.
Let the wavenumbers satisfy $k_1\ll k_2\simeq k_3$. 
In this 
limit, the contributions of two IR regions which include $k_1$ are dominant
(see Fig.~\ref{divergence of bi}). 
Therefore {(\ref{bis:SSS:int}) can be written as
\begin{eqnarray}\label{BSSS:pow}
	B_{\mathcal S\mathcal S\mathcal S}(k_1,k_2,k_3)&\simeq&
	\frac{8}{n_g-1}\gamma^{3}_{2}\mathcal P
	_g(k_0)
\left[\left(\frac{k}{k_0}\right)^{n_g-1}\right]^{k_{\rm max}}_{L^{-1}}
	 \nonumber \\
	&&\times\left[P_g(k_1)P_g(k_2)+P_g(k_1)P_g(k_3)\right], \quad\text{for $k_1\ll k_2,k_3$,}
\end{eqnarray}
where $k_\mathrm{max}\ltsim k_1$ and we will use $k_\mathrm{max}=k_1$ hereafter.
Here we also introduce IR cut-off to be the current horizon size $\sim L$.

\begin{figure}
	\begin{center}
		\includegraphics[width=8cm]{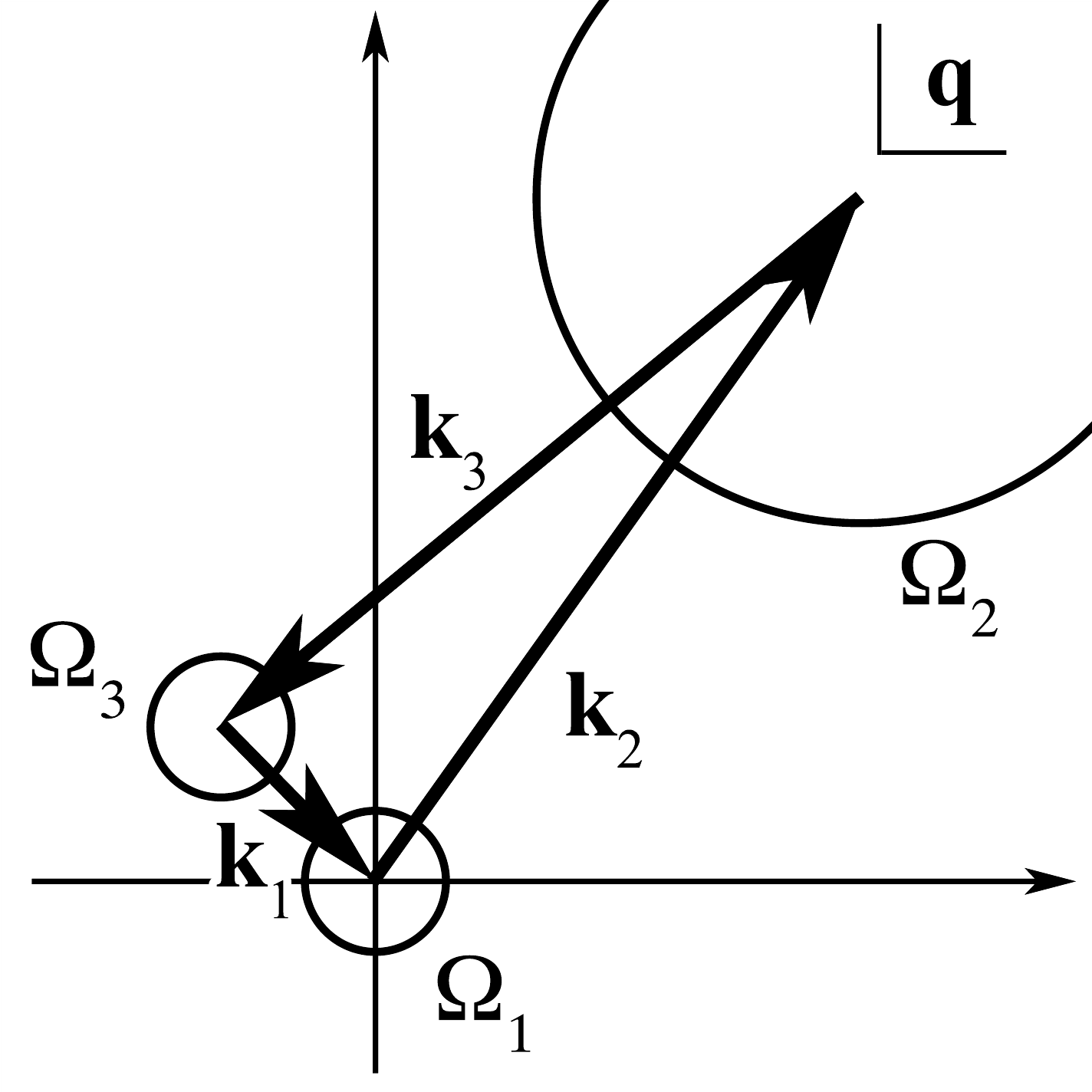}
		\caption{The IR regions of the integration for the bisectrum in the squeezed limit ($k_1\ll k_2,\,k_3$).
		Since $P_g(k_1)\gg P_g(k_2),\,P_g(k_3)$ in this case, the contributions from $\Omega_1$ and $\Omega_{3}$,
		which are approximately proportional to $P_g(k_1)P_g(k_2)$ and $P_g(k_1)P_g(k_3)$, are
		much larger than that from $\Omega_{2}$, which are approximately proportional to $P_g(k_2)P_g(k_3)$.
		Therefore we can consider the expansions of power spectrum 
		only in $\Omega_1$ and $\Omega_3$. At this time, the upper limit of these IR regions is around $k_1/2$.
		From now on, we take $k_\mathrm{max}=k_1$.
		}
		\label{divergence of bi}
	\end{center}
\end{figure}

\section{CMB $\mu$-distortion}
\label{CMB mu distortion}
\subsection{Homogeneous distortions}
\label{Homogeneous distortions}

We usually assume that the photon system is locally thermal equilibrium in the early universe.
In other words, the photon fluid has generally local Bose-Einstein distribution function defined as
\begin{align}
f({\bm x},\omega)=\frac{1}{e^{\frac{\omega }{T_\text{BE}({\bm x})}+\mu({\bm x})}-1}\label{36},
\end{align}
where $\bm x$ and $\omega$ are spacetime point and frequency, respectively.
Note that the temperature parameter of the Bose-Einstein distribution function $T_{\rm BE}$ is different from that of Planck distribution function $T_{\rm pl}$. 
Assuming that the deviations from the Planck distribution function are induced by the mixing of different local blackbodies,
$T_{\rm BE}- T_{\rm pl}$ and $\mu$ are the second-order quantities of the first-order dimensionless temperature perturbation defined as~\cite{Khatri:2012rt}
\begin{align}
\Theta(\bm x)=\frac{T_{\rm pl}(\bm x)-\langle  T_{\rm pl}\rangle }{\langle  T_{\rm pl}\rangle}.
\end{align}
Then, using
the conservation laws of both average energy and number, we can derive the evolution equation for the averaged $\mu$-distortion as follows~\cite{Ota:2014hha}
\begin{align}
\frac{d}{d\eta} \langle \mu \rangle =-\frac{\langle \mu\rangle }{t_\mu}-1.4 \times
 4\left\langle \Theta \frac{d\Theta}{d\eta}\right\rangle +\mathcal O(\Theta^3),
\label{eq:diff_mu2}
\end{align}
where $\eta$ is conformal time and 
the first term is added by hand to take into account the effect of the process which do not conserve the number of photon such as the double Compton effect or 
electron-positron pair annihilation.
$t_\mu$ is
the time-scale of decreasing of the chemical potential by the above processes, which become ineffective at
$z\lesssim2\times 10^6$~\cite{Gould:1984,Hu:1992dc}.
$d\Theta /d\eta$ is immediately calculated by linear Boltzmann equations.
Here let us follow the notation in Ma and Bertschinger~\cite{Ma:1995ey}. 
In Fourier space, the brightness functions for intensity and linear poralization are given by
\begin{align}
F_{\gamma}&=\frac{\int q^2dq qf^{(0)}(q)\Psi}{\int q^2dq
 qf^{(0)}(q)},
\label{eq:def_fg}
\\ 
G_{\gamma}&=\frac{\int q^2dq qf^{(0)}(q)\Psi_P}{\int q^2dq qf^{(0)}(q)},
\label{eq:def_gg}
\end{align}
where $q$ is comoving momentum and $f^{(0)}$ is the background Planck distribution function.
$\Psi$ and $\Psi_P$ are fractional perturbations in
$\rho_{11}+\rho_{22}$ and $\rho_{11}-\rho_{22}$ of the photon density matrix, respectively. 
We expand these quantities with respect to multipoles as $F_{\gamma}=\sum_{l=0}(-i)^l(2l+1)P_l(\lambda)F_{\gamma l}$ to obtain the solutions order by order.
To the linear order, $F_\gamma = 4\Theta$ is always satisfied.
Then we can derive the following formulae for the $\mu$-distortion~\cite{Chluba:2012gq}
\begin{align}
\langle\mu\rangle=&1.4\cdot \frac14\int^{\eta_{f}}_{0}d\eta'\mathcal J_{DC}(\eta')
\int d(\ln k) \left[\mathcal P_G(k)+\left(f+\frac{\gamma_1}{\gamma_2}
\right)^2\mathcal P_{\mathcal S \mathcal S}(k)\right]
\notag \\ &\times n_e\sigma_Ta \bigg[
\frac{3}{4}(F_{\gamma1}-F_{b1})^2-\frac{F_{\gamma2}}{2}(-9F_{\gamma 2}+G_{\gamma2}+G_{\gamma0})+\sum_{l\ge3}
(2l+1)F_{\gamma l}F_{\gamma l}\bigg], \label{mus}
\end{align}
where $\mathcal J_{DC}$ is a window function induced by double Compton scattering and $\eta_f$ is the end time of $\mu$ era.
$F_{b1}$ is the velocity perturbation of baryons.
Here, $f=-R_\nu/(4 R_\gamma)$ 
with $R_\nu=\rho_\nu/(\rho_\nu +\rho_\gamma)$ and $ R_\gamma=1-R_\nu$ is the ratio of the NID mode to the adiabatic mode and the transfer functions have the same form with (\ref{FFtransfer})~\cite{Bucher:1999re}.
We replace $F_{\gamma1} \to F_{\gamma1} -F_{b1}$ to manifest gauge invariance of the first term,
which can be omitted in the radiation dominated period.
Note that Legendre coefficients and $f$ only depend on the magnitude of $\bm k$.
In the tight coupling regime, we can approximately 
solve Boltzmann equations for the photon sector analytically.
In the case with adiabatic condition, the solution is given by~\cite{Dodelson:2003ft}
\begin{align}
F_{\gamma1}\sim -\frac{4}{\sqrt{3}}\sin(kr_s)\exp\left(-\frac{k^2}{k_D^2}\right),
\label{eq:transf}
\end{align}
where $r_s$ is the sound horizon and $k_D$ is the Silk damping scale.
In the tight coupling regime we can write $G_{\gamma 0}+G_{\gamma 2}=3F_{\gamma 2}/2$, then (\ref{mus}) can be approximate as
\begin{align}
\langle\mu\rangle&= -2.8\int^{\eta_{f}}_{0}d\eta\mathcal J_{DC}(\eta)
\int d(\ln k) \left[\mathcal P_G(k)+\left(f+\frac{\gamma_1}{\gamma_2}
\right)^2\mathcal P_{\mathcal S \mathcal S}(k)\right]
\partial_\eta \exp\left(-2\frac{k^2}{k_D^2}\right)\notag\\
&\sim -2.8
\int d(\ln k)\left[\mathcal P_G(k)+\left(f+\frac{\gamma_1}{\gamma_2}
\right)^2\mathcal P_{\mathcal S \mathcal S}(k)\right]
\bigg[\exp\left(-2\frac{k^2}{k_D^2}\right)\bigg]^f_i,\label{mudisanal}
\end{align}
where we have used the relation $F_{\gamma2} = 8kF_{\gamma1}/(15 \dot\tau)$ and replace 
$\sin^2(kr_s)$ with 1/2, and in addition we have adopted
the relation $\partial_\eta k_D^{-2}=-8/(45\dot\tau)$ in the limit of full radiation domination. 
Let us divide $\langle \mu \rangle $ into parts originating from $\mathcal R_G$ and $g^2$,
that is, $\langle\mu\rangle =\langle \mu \rangle_G+\langle \mu \rangle_{g^2}$.
Then (\ref{mudisanal}) yields
\begin{align}
\langle \mu \rangle_G 
&= 2.8\mathcal P_G(k_0)\log\left(\frac{k_{Di}}{k_{Df}}\right)\left[1+\frac{n_s-1}{2}\left(\log\left(\frac{k_{Di}k_{Df}}{2k_0^2} \right)-{\bf C}\right)+\cdots \right]
\notag \\
&= 3.36\times 10^{-8}[1+8.91(n_s-1)+\cdots ],
\end{align}
where $\mathbf C=0.577...$ is Euler-Mascheroni constant and $\log(k_{Di}/k_{Df})\simeq 5.477$. 
If we take into account terms of $\mathcal O(n_s-1)$ with $n_s=0.96$,
$\langle \mu \rangle$ becomes $2.2\times 10^{-8}$, which is consistent with the values derived in the previous works~\cite{Hu:1994bz,Sunyaev:1970eu,
1991MNRAS.248...52B,1991ApJ...371...14D}.
The second order correction is smaller than 10\%.
Another contribution from $g^2$ can be calculated as
\begin{align}
\langle \mu \rangle_{g^2} 
=&1.4\mathcal P_{\mathcal S \mathcal S}(k_0)\left(\frac{R_\nu}{4 R_\gamma}+\frac{\gamma_1}{\gamma_2}\right)^2
\left[\left(\frac{k_{D}}{\sqrt{2}k_0}\right)^{2n_g-2}\right]^i_f\Gamma(n_g-1)\notag \\
=&\frac{5.6\gamma^{2}_{2}\mathcal P^2_g(k_0)}{n_g-1}
\left(\frac{R_\nu}{4 R_\gamma}+\frac{\gamma_1}{\gamma_2}\right)^2\left[\left(\frac{k}{k_0}\right)^{n_g-1}\right]^{k_{\rm max}}_{L^{-1}} \left[\left(\frac{k_{D}}{\sqrt{2}k_0}\right)^{2n_g-2}\right]^i_f\Gamma(n_g-1),\label{mudis:g2}
\end{align}
where we have used (\ref{pow:ss:cutoff}).
Suppose that uncorrelated case with $\gamma_1=0$ and $\gamma_2=1$, expanding around the flat case, we obtain	
\begin{align}
\langle\mu\rangle^{\rm uncor}_{g^2}\simeq12.1
\mathcal P^2_{g}(k_0)\bigg[1. + 13.9 (n_g-1)+ \cdots \bigg],\label{unc:mu:g2origin:expand}
\end{align}
where $k_{\rm max}= k_0$.
Fig.~\ref{g2_iso_mu1} shows $\langle \mu\rangle^{\rm uncor}_{g^2}$ 
as a function of $n_g$.
For $\mathcal{S}$ not to dominate $\mathcal{R}$, we should impose $\mathcal{P}_\mathcal{SS}(k_0)\sim\mathcal{P}_g^2(k_0)\ltsim10^{-10}$
(we obtain similar constraints from the CMB angular bispectrum and see Appendix~\ref{app2} for details). For example, assuming 
$\mathcal{P}_g(k_0)\sim10^{-5}$, it can be shown from (\ref{unc:mu:g2origin:expand})
that $\braket{\mu}_{g^2}^\mathrm{uncor}\sim10^{-9}$ for 
the case of flat spectrum $n_g\sim1$.
As shown in Fig.~\ref{g2_iso_mu1}, if we consider the blue-tilted power spectrum of $g$,
we can realize the large enhancement of $\braket{\mu}$. From this figure, by employing the COBE FIRAS constraint, that is, 
$\mu < 9 \times 10^{-5}$, we find an upper limit on $n_g$ as $n_g \lesssim 1.5$.

\begin{figure}
\centering
\includegraphics[width=12cm]{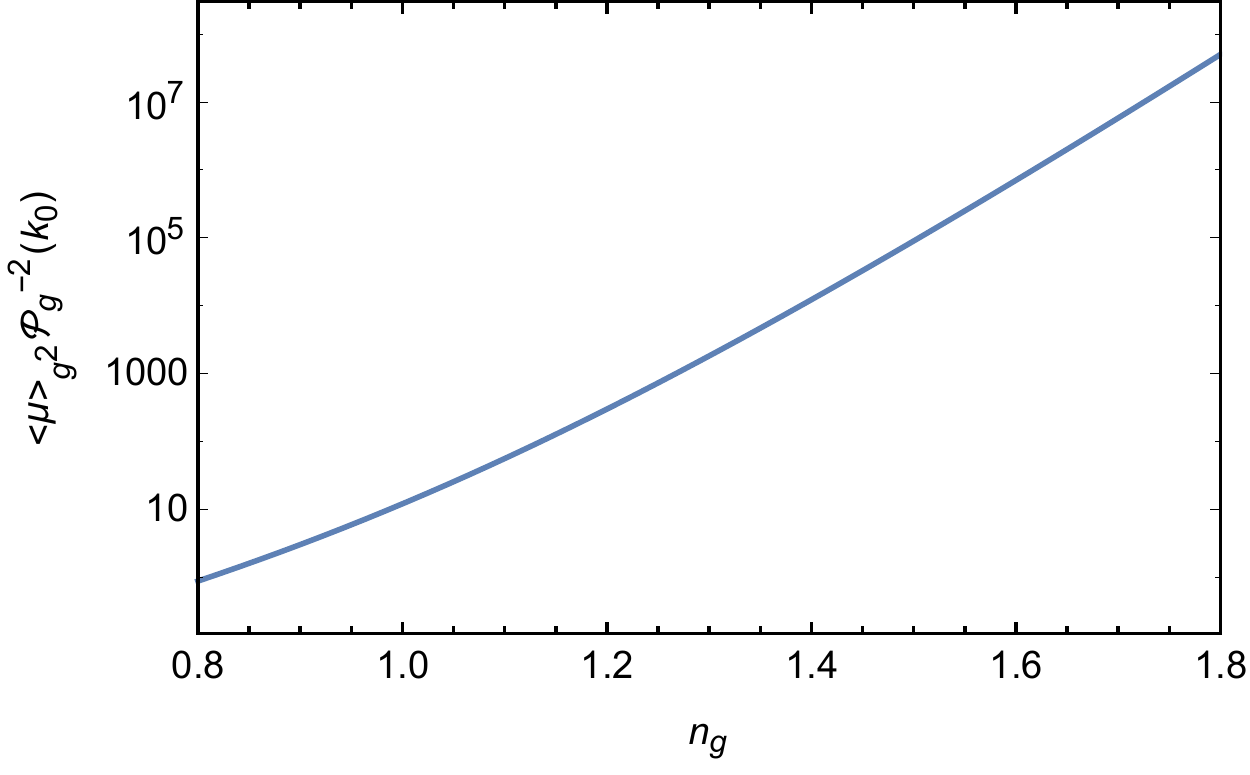}
\caption{$n_g$ v.s. $\langle \mu\rangle^{\rm uncor}_{g^2}$ in units of $\mathcal P^2_g(k_0)$.
Here $\gamma_1$ and $\gamma_2$ are set to zero and unity, respectively.
}
\label{g2_iso_mu1}
\end{figure}

Incidentally,
the $\mu$-distortion originating from almost Gaussian uncorrelated NID 
mode is also calculated straightforwardly
and the difference from the case of the adiabatic perturbation is only from that of the transfer functions of them.
Specifically, the $\mu$-distortion from the isocurvature mode becomes
$\mathcal O(0.01)$ times smaller than that of the standard adiabatic perturbation case (see \cite{Chluba:2013dna} for more details).

\subsection{Inhomogeneous distortions}

In the case with inhomogeneous distortions, (\ref{eq:diff_mu2}) does not work since the equation is a result of conservation laws in the homogeneous and isotropic background.
In other words, the quantities in the equation are globally averaged.
Therefore, generally we should consider local conservation laws of energy-momentum tensor
$\nabla_\mu T^{\mu \nu}=0$ and number flux $\nabla_\mu N^\mu=0$
to compute inhomogeneous distortions~\cite{Pajer:2013oca}.
Nevertheless approximately we can just remove $\langle\cdots \rangle$ of (\ref{eq:diff_mu2}) since the thermodynamic valuables are locally averaged quantities.
Therefore we again use (\ref{mudisanal}) with the slight change
\begin{align}
\int d(\ln k) \mathcal P_{\mathcal X\mathcal Y}(k)\to
\int \frac{d^3k}{(2\pi)^3}\frac{d^3k'}{(2\pi)^3} e^{i(\bm k+\bm k')\cdot \bm x}\mathcal X(\bm k)\mathcal Y(\bm k')5P_2(\hat k\cdot \bm n)P_2(\hat k'\cdot \bm n),\mbox{~with~}\mathcal X,\mathcal Y=\mathcal R, \mathcal S,
\end{align}
where $\hat k$ and $\hat k'$ are unit vector of $\bm k$ and $\bm k'$.
$\bm n$ is a tangent vector of the line-of-sight of the photons.
Then we obtain the $\mu $ distortion in Fourier space as follows:
\begin{align}
\mu(\eta_{f},\bm k)= 16\alpha \int \frac{d^3k_1}{(2\pi)^3}\left[\mathcal R_{G\bm k_1}\mathcal R_{G\bm k_2}+
\left(
\frac{R_\nu}{4 R_\gamma}+
\frac{\gamma_1}{\gamma_2}
\right)^2\mathcal S_{\bm k_1} \mathcal S_{\bm k_2}
\right]\notag \\
\times5P_2(\hat k_1\cdot \bm n)P_2(\hat k_2\cdot \bm n) \langle\sin k_1r_s\sin k_2r_s\rangle _p\left[\exp\left(-\frac{k_1^2+k_2^2}{k_D^2}\right)\right]^i_f,\label{ft:of:mu}
\end{align}
where
$\alpha= -1.4\times1/4$, 
$\bm k_2=\bm k-\bm k_1$ and $\langle \cdots \rangle_p$ is a periodic average.

\section{$\mu T$ angular cross-correlation}
\label{muT angular cross correlation}

The harmonic coefficients of observed anisotropies in $\Theta$ and $\mu$ can be written as
\begin{align}
a_{T,lm}=&\int d\bm nY^*_{lm}(\bm n)\Theta(\eta_0,\bm x,\bm  n),
\label{eq:alm}\\
a_{\mu,lm}=&\int d\bm nY^*_{lm}(\bm n) \mu(\eta_0,\bm x,\bm  n),\label{eq:alm:mu}
\end{align}
where $\eta_0$ is the conformal time today and $\bm n$ is a line-of-sight.
Without loss of generality, we can take $\bm x=0$.
Then the angular power spectrum of the $\mu T$ cross-correlation function is
defined as usual, 
\begin{align}
C^{\mu T}_l=\frac{1}{2l+1}\sum_m\langle a^*_{\mu,lm}a_{T,lm}\rangle.
\end{align}
In Fourier space, the temperature perturbations are 
\begin{equation}
\Theta(\eta,\bm k,\bm n)=
\Theta^{\mathcal R}(\eta; k,\lambda)\mathcal R_{\bm k}+\Theta^{\mathcal S}(\eta; k,\lambda)\mathcal S_{\bm k},
\label{Theta:legendre:ex}
\end{equation}
where
$\lambda$ is the cosine between $\bm k$ and $\bm n$.
On the other hand, the $\mu$-distortion is written as
\begin{align}
\mu(\eta,\bm k,\bm n)=\Delta_\mu(\eta;\eta_f,k,\lambda)\mu(\eta_f,\bm k),
\end{align}
where $\mu(\eta_f,\bm k)$ is what we calculated in the last section and the transfer function is given in Ref.~\cite{Pajer:2013oca}.
Substituting (\ref{Theta:legendre:ex}) and (\ref{ft:of:mu}) into (\ref{eq:alm}) and (\ref{eq:alm:mu}),
we obtain
\begin{align}
a_{T,lm}&=4\pi(-i)^l\int \frac{d^3k}{(2\pi)^3}Y^*_{lm}(\hat k) \left[\Theta^{\mathcal R}_{l}(\eta_0,k)\mathcal R_{G \bm  k}+\left(\Theta^{\mathcal S}_{l}(\eta_0, k)+\frac{\gamma_1}{\gamma_2}\Theta^{\mathcal R}_{l}(\eta_0,k)\right)\mathcal S_{\bm k}\right],\\
a_{\mu,lm}&=4\pi(-i)^l\cdot 16\alpha \int  \frac{d^3k}{(2\pi)^3} \frac{d^3k_1}{(2\pi)^3}Y^{*}_{lm}(\hat k)\Delta_{\mu l}(\eta_0, k)\langle\sin k_1r\sin k_2r\rangle _p 5P_2(\hat k_1\cdot \bm n)P_2(\hat k_2\cdot \bm n)
\notag \\
&
\times\left[\exp\left(-\frac{k_1^2+k_2^2}{k_D^2}\right)\right]^i_f
\left[\mathcal R_{G\bm k_1}\mathcal R_{G\bm k_2}+
\left(
\frac{R_\nu}{4 R_\gamma}+
\frac{\gamma_1}{\gamma_2}
\right)^2\mathcal S_{\bm k_1} \mathcal S_{\bm k_2}
\right]
,
\end{align}
where we define $\hat k=\bm k/k$, and 
$\Theta^X_l(X=\mathcal R,\mathcal S)$ and $\Delta_{\mu l}$ are the Legendre coefficients of the transfer functions.
On large angular scales where the Sachs-Wolfe effect is dominant, 
$\Theta_l$'s are given by
\begin{align}\label{eq:SW}
\Theta_{l}(\eta_0,k)\sim \left[\Theta_{0}(\eta_*)+\psi(\eta_*)\right]j_l(k(\eta_0-\eta_*)),
\end{align}
where $\eta_*$
is conformal time at recombination and $\psi$ is the gravitational potential in the conformal Newtonian gauge.
$\Theta_{0}(\eta_*)+\psi(\eta_*)$ depends on the initial conditions. 
Using the numerical code CLASS~\cite{Blas:2011rf}, 
$\Theta_{0}(\eta_*)+\psi(\eta_*)\sim -0.24$ for the adiabatic perturbation 
 and $\Theta_{0}(\eta_*)+\psi(\eta_*)\sim -0.175$ for the neutrino isocurvature density mode, respectively. 
$\Delta_{\mu l}$ is also given by the line-of-sight integral method as shown in Ref.~\cite{Pajer:2013oca}.
Using the above equations, $\mu T$ angular power spectrum is obtained as
\begin{align}
|C^{\mu T}_l|=
\frac{(4\pi)^2 \alpha}{2l+1} \left(
0.175+0.24\frac{\gamma_1}{\gamma_2}
\right)\left(\frac{R_\nu}{ R_\gamma}+\frac{4\gamma_1}{\gamma_2}\right)^2
\int \frac{d^3kd^3k_1d^3k_2}{(2\pi)^9}\sum^l_{m=-l}Y^*_{lm}(\hat k)Y_{lm}\left(\frac{\bm k_1+\bm k_2}{|\bm k_1+\bm k_2|}\right)
\notag \\
\times5P_2(\hat k_1\cdot \bm n)P_2(\hat k_2\cdot \bm n)\langle \mathcal S_{\bm k}\mathcal S^*_{\bm k_1}\mathcal S^*_{\bm k_2}\rangle j_l(k\eta_0)j_l(|\bm{k}_1+\bm{k}_2|\eta_0)\langle\sin k_1r_s\sin k_2r_s\rangle _p\left[\exp\left(-\frac{k_1^2+k_2^2}{k_D^2}\right)\right]^i_f,
\end{align}
where $r_s$ is the sound horizon and $\eta_0\gg \eta_*$. 
The three-point function $\langle \mathcal{SSS}\rangle$ is nonzero since $\mathcal S$ is non-Gaussian.
From the reality
condition, we obtain 
\begin{align}
\langle \mathcal S_{\bm k}\mathcal S^*_{\bm  k_1}\mathcal S^*_{\bm k_2}\rangle=\langle \mathcal S_{\bm k}\mathcal S_{-\bm k_1}\mathcal S_{-\bm k_2}\rangle=(2\pi)^3\delta^{(3)}(\bm k -\bm  k_1-\bm  k_2)B_{\mathcal S\mathcal S\mathcal S}(k,k_1,k_2).
\end{align}
Let us consider a
transformation $\bm k_\pm=\bm k_1\pm \bm k_2$.
Since we will concentrate on low-$l$, the integrations of $\bm{k}$ and $\bm{k}_+$ are negligible except
around CMB scales $k=k_+\sim k_0$,
because of the behaviour of the spherical Bessel functions. On the other hand, due to the exponential suppression factor
the contribution around $k_1^2+k_2^2\sim k_D^2\gg k_0$
is dominant. Therefore, from $k_+^2+k_-^2=2(k_1^2+k_2^2)\gg k_+^2$, we obtain a hierarchical relation $k_-\gg k_+$. At this time, since 
$k_1\sim k_2\sim k_-/2$, we can approximate the periodic average $\braket{\sin k_1r\sin k_2r}_p$ by $1/2$. 
From the above results,
noting that the Jacobian of coordinate transformation to $\bm k_\pm$ is 1/8 and we consider the squeezed configuration now,
we obtain

\begin{align}
&B_{\mathcal S\mathcal S\mathcal S}(k_+,k_-/2,k_-/2)\notag \\
&\sim  \frac{2\pi^2}{k_+^3}\frac{2\pi^2}{(k_-/2)^3}
\left(\frac{k_+}{k_0}\right)^{n_g-1}\left(\frac{k_-/2}{k_0}\right)^{n_g-1}
 \frac{16\mathcal P^3_g(k_0)}{n_g-1}
\left[\left(\frac{k}{k_0}\right)^{n_g-1}\right]^{k_{\rm max}}_{L^{-1}}
,
\end{align}
with use of (\ref{BSSS:pow}).
Finally we have the following form
\begin{align}
|C^{\mu T}_l|
&\sim 0.583\left(1+1.4\frac{\gamma_1}{\gamma_2}\right)\left(1+5.8\frac{\gamma_1}{\gamma_2}\right)^2 \frac{\gamma^3_2\mathcal P^3_g(k_0)}{n_g-1}
\left[\left(\frac{k}{k_0}\right)^{n_g-1}\right]^{k_{\rm max}}_{L^{-1}}
\notag \\
&\times\left[\left(\frac{\sqrt 2 k_D}{k_0^2L}\right)^{n_g-1}\right]^i_f\frac{\Gamma\left(l+\frac {n_g}2-\frac12\right)\Gamma(3-n_g)\Gamma\left(\frac{n_g-1}{2}\right)}{\Gamma\left(l+\frac 52-\frac {n_g}2\right)\Gamma^2\left(2-\frac {n_g}2\right)}
,
\end{align}
where we have used $P_2(-\hat k_- \cdot \bm n)=P_2(\hat k_- \cdot \bm n)$ in the integration with respect to $\hat k_-$ and $\eta_0=L$.
For the flat and uncorrelated spectrum with $\mathcal P_g\sim 10^{-5}$, which would be marginally
allowed by CMB bispectrum from Planck (See Appendix~\ref{app2} for the Planck forecast), 
and taking $k_{\rm max}=k_0$, we obtain $|l(l+1)C^{\mu T}_{l}|\sim 10^{-14}$.
This level of signal corresponds to $f^{\rm loc}_{\rm NL}\sim 100$ in the case of the local-type non-Gaussianity in adiabatic perturbations~\cite{Pajer:2012vz}, 
which is 10 times smaller than the expected sensitivity
of PIXIE and comparable to that of PRISM.
For $l=10$ with $k_{\rm max}=k_0$ and $\gamma_1=0$, 
\begin{align}
\frac{(10\cdot 11)\times| C^{\mu T}_{10}|}{\mathcal P^3_g(k_0)}\sim 53.4\bigg[
1. + 1.75514 (n_g-1) +\cdots \bigg].
\end{align}
Fig.~\ref{g2_iso_muT} shows that $n_g$-dependence of $110C^{\mu T}_{10} \mathcal P_g^{-3}(k_0)$.
In the case with $n_g\sim 1.5$, the signal is almost 10 times bigger than that with the flat case, and can be also detected by PIXIE.
\begin{figure}
\centering
\includegraphics[width=13cm]{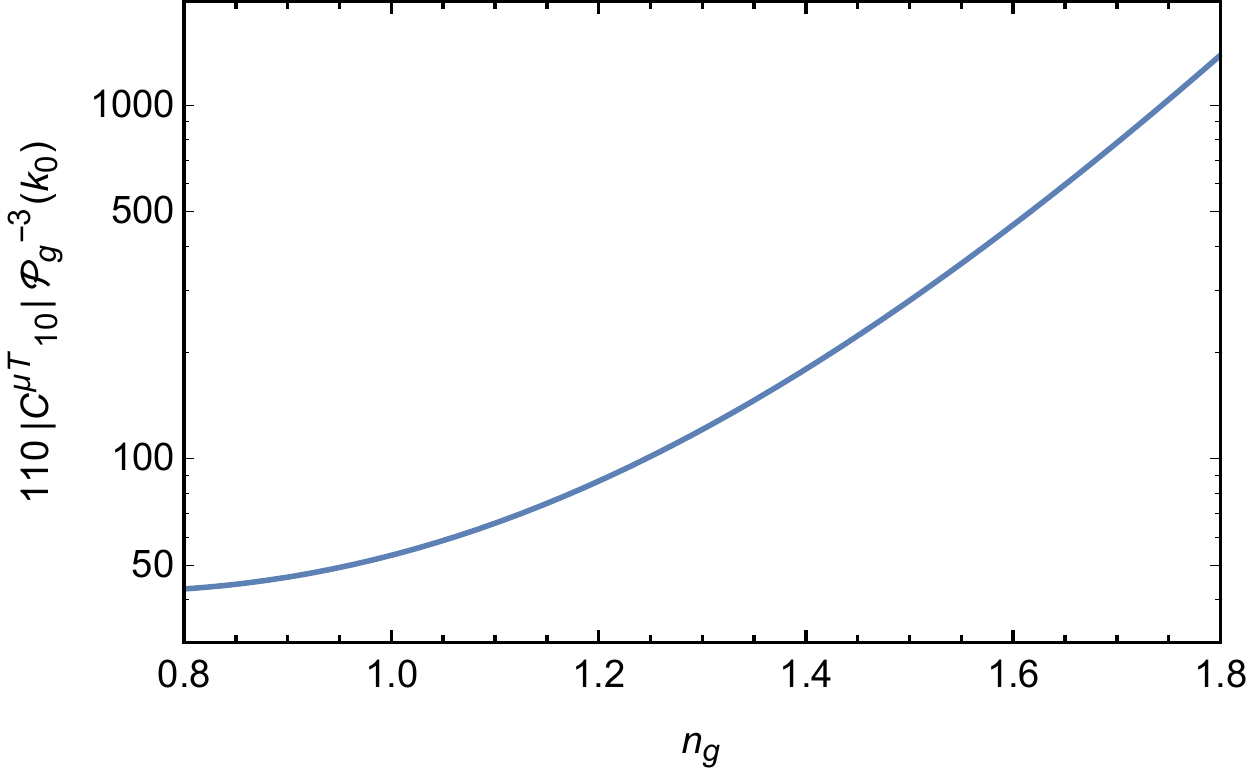}
\caption{$n_g$ dependence of $l(l+1)C^{\mu T}_{l} \mathcal P_g^{-3}(k_0)$ with $l=10$.}
\label{g2_iso_muT}
\end{figure}

For the sake of completeness, let us consider the case with almost Gaussian neutrino isocurvature density mode,
$\mathcal{S}=\mathcal{S}_G+f_\mathrm{NL}^{\mathrm{loc},\nu}(\mathcal{S}_G^2-\braket{\mathcal{S}_G^2})$.
Local-type bispectrum of the uncorrelated isocurvature perturbation is parametrized as
\begin{align} \label{loc_nu}
B_{\mathcal S}(k_+,k_-/2,k_-/2)=-\frac65f^{{\rm loc},\nu}_{\rm NL}[P_{\mathcal S}(k_+)P_{\mathcal S}(k_-/2)+2{\rm perm.}]\sim 
-\frac{6\times 2}5f^{{\rm loc},\nu}_{\rm NL}P_{\mathcal S}(k_+)P_{\mathcal S}(k_-/2).
\end{align}
As discussed below \eqref{eq:transf} and \eqref{eq:SW}, transfer functions of the adiabatic and NID modes differ
only by a multiplicative constant factor. Thus by taking account this difference in the results of Ref.~\cite{Pajer:2012vz}, we obtain
\begin{align}
l(l+1)C^{\mu T}_l
&\lesssim 8.04\times 10^{-19}f^{{\rm loc},\nu}_{\rm NL}\left[1+\left(5.37+
\frac{1}{l}+\frac{1}{l+1}+2\psi^{(0)}(l)
\right)\frac{n_s-1}{2}+\cdots \right],
\end{align}
where we have assumed $\beta_{\rm NID} = 0.27$ and
$\psi^{(0)}(x)=d\log\Gamma(x)/dx$ is a poly-gamma function.
When $f_{\rm NL}=f^{{\rm loc}, \nu}_{\rm NL}$, the $\mu T$ cross-correlation from the 
non-Gaussian neutrino isocurvature perturbations should be 
$10^2$ times smaller than that from the adiabatic ones calculated in Ref.~\cite{Pajer:2012vz}.
However as is shown in Appendix~\ref{app2}, the expected observational constraints on $f^{{\rm loc}, \nu}_{\rm NL}$ would be 
$10^4$ at around 2$\sigma$ level when $\beta_{\rm NID}\simeq \mathcal P_\mathcal{SS}/\mathcal P_\mathcal{RR}=10^{-1}$.
When we take $f_\mathrm{NL}^{\mathrm{loc},\nu}$ to be $10^4$,
the cross-correlation from the non-Gaussian neutrino isocurvature perturbations can be as large as that from local-type 
adiabatic ones with $f_{\rm NL}=100$. 
The size of signal here is the same as in the Gaussian-squared case, which we have presented before.
This is by no means surprising since both the amplitudes of $\mu T$ cross-correlation and the CMB bispectrum are determined 
by the primordial bispectrum, whose spectral shape can be approximated with the local-type one both in the cases 
of weakly non-Gaussian and Gaussian-squared isocurvature perturbations.

\section{Conclusions}

In this paper, we have calculated the mean $\mu$-distortion and the 
cross-correlation of its anisotropy with primary CMB temperature one
in the presence of non-Gaussian neutrino isocurvature perturbations.
In particular, 
we have focused on 
the Gaussian-squared
perturbations,
and explicitly shown that the primordial bispectrum of Gaussian-squared type perturbations
can be approximated by the local-type one.
We have found that when the power spectrum of the isocurvature perturbations are nearly scale-invariant, 
the mean $\mu$ and the $\mu T$ cross-correlation can be as large as $10^{-9}$ and $10^{-14}$
with the present constraints from CMB power spectrum and bispectrum on NID mode being satisfied.
In particular, $\mu T$ cross-correlation from NID perturbations is potentially observed by the PRISM surveys, which
is contrastive to the cases of adiabatic local-type ones, which requires $f_{\rm NL}$ an order of magnitude larger than the
upper bound from current CMB bispectrum measurements. 
If the power spectrum of isocurvature perturbations are allowed to be blue-tilted, 
the $\mu T$ cross-correlation can be enhanced by an order of magnitude and expected to be observed by PIXIE.

\acknowledgments 
We would like to thank Masahide Yamaguchi for the helpful discussions and comments.
This work was supported by World Premier International Research Center Initiative (WPI Initiative), MEXT, Japan.
Y.T. is supported by an Advanced Leading Graduate Course for Photon Science grant.
T.S. is supported by the Academy of Finland grant 1263714.

\appendix

\section{Constraints from CMB angular bispectrum} \label{app2}
In this appendix, we summarize the Fisher matrix analysis of CMB bispectrum and 
forecast for the Planck constraints on non-Gaussian isocurvature perturbations, 
following Refs.~\cite{Langlois:2011hn,Kawakami:2012ke,Langlois:2012tm}\footnote{
We defer it for future work to derive constraints on non-Gaussian 
neutrino isocurvature perturbations from the actual Planck data here.
Constraints from the WMAP data can be found in Ref.~\cite{Hikage:2012tf}.}.
Let us start by generalizing the form of CMB anisotropies in~\eqref{eq:alm} into
\begin{equation}
a^P_{lm}=4\pi (-i)^l
\int \frac{d^3k}{(2\pi)^3}\sum_X
g^{XP}_l(k) Y_{lm}^*(\hat k)\mathcal X^X(\bm k), 
\end{equation}
where $\hat k$ is the unit vector of $\bm k$, $X$ and $P$ respectively represent types of initial perturbations
and CMB anisotropies, i.e. $X$ is either the adiabatic ($\mathcal R$) or neutrino isocurvature ($\mathcal S$) mode and $P$ is either 
the temperature (T) or E-mode polarization (E) anisotropy.
$g^{XP}_{l}$ is the Legendre coefficient of the transfer function of $P$ from $X$
 and numerically evaluated using the {\tt CAMB} code~\cite{Lewis:1999bs}.
The CMB bispectrum in harmonic space is given as 
\begin{equation}
B^{P_1P_2P_3}_{l_1m_1l_2m_2l_3m_3} \equiv \langle a^{P_1}_{l_1m_1}
a^{P_2}_{l_2m_2}a^{P_3}_{l_3m_3}\rangle.
\label{eq:bi1}
\end{equation}
Given a primordial bispectrum
\begin{equation}
\langle \mathcal X^{X_1}(\bm k_1)\mathcal X^{X_2}(\bm k_2)\mathcal X^{X_3}(\bm k_3)\rangle
=B^{X_1X_2X_3}(k_1,k_2,k_3)(2\pi)^3\delta^{(3)}(\bm k_1+\bm k_2+\bm k_3),
\end{equation}
\eqref{eq:bi1} can be rewritten as 
\begin{eqnarray}
B^{P_1P_2P_3}_{l_1m_1l_2m_2l_3m_3}
&=&
\sum_{X_1X_2X_3}
\prod^3_{i=1}\left[
4\pi (-i)^{l_i}
\int\frac{d^3k_i}{(2\pi)^3}
g^{X_iP_i}_{l_i}(k_i)Y^*_{l_im_i}(\hat k_i)
\right] \notag \\
&&\quad\times
B^{X_1X_2X_3}(k_1,k_2,k_3)(2\pi)^3
\delta^{(3)}(\bm k_1+\bm k_2+\bm k_3).
\label{eq:bi2}
\end{eqnarray}
By Fourier transforming the delta function 
and using a formula for the partial wave decomposition 
\begin{equation}
e^{i\bm k\cdot \bm r}=\sum_{lm} 4\pi i^l j_l(kr) Y_{lm}(\hat k) Y^*_{lm}(\hat r), 
\end{equation}
\eqref{eq:bi2} can be further recast into
\begin{equation}
B^{P_1P_2P_3}_{l_1m_1l_2m_2l_3m_3}
=\mathcal G^{m_1m_2m_3}_{l_1l_2l_3}
b^{P_1P_2P_3}_{l_1m_1l_2m_2l_3m_3}, 
\label{eq:rb1}
\end{equation}
where 
\begin{equation}
b^{P_1P_2P_3}_{l_1m_1l_2m_2l_3m_3}=\sum_{X_1X_2X_3}
\int r^2dr\,
\prod_{i=1}^3\left[\frac{2}{\pi}\int k_i^2dk_i
\,g^{X_iP_i}_{l_i}(k_i)j_{l_i}(k_ir)
\right] B^{X_1X_2X_3}(k_1,k_2,k_3)
\end{equation}
is the reduced bispectrum and 
\begin{equation}
\mathcal G^{m_1m_2m_3}_{l_1l_2l_3}=
\int d\hat r \prod^3_{i=1}Y^*_{l_im_i}(\hat r)
\end{equation}
is the Gaunt integral, which can be represented in terms 
of the Wigner-3j symbol as 
\begin{equation}
\mathcal G^{m_1m_2m_3}_{l_1l_2l_3}
=\sqrt{\frac{(2l_1+1)(2l_2+1)(2l_3+1)}{4\pi}}
\begin{pmatrix}
l_1 & l_2 & l_3 \\
0 & 0 & 0 
\end{pmatrix}
\begin{pmatrix}
l_1 & l_2 & l_3 \\
m_1 & m_2 & m_3 \\
\end{pmatrix}.
\end{equation}

Now let us consider the non-Gaussian perturbations 
given in \eqref{eq:defR}-\eqref{eq:defS}.  As is shown in \eqref{BSSS:pow}, 
the bispectrum of these Gaussian-squared type perturbations
can be approximated in the form of the local-type bispectrum, 
\begin{equation}
B^{X_1X_2X_3} (k_1,k_2,k_3)\simeq 
8A \mathcal 
\gamma_{X_1}\gamma_{X_2}\gamma_{X_3}
\mathcal P_g(k_0) \left[
P_g(k_1)P_g(k_2)+(\mbox{2 cyclic perms})\right],
\end{equation}
where the factor $\gamma_X$ should be $\gamma_1$ for $X=\mathcal R$
and $\gamma_2$ for $X=\mathcal S$. Here, the factor $A\simeq 
\left[(k/k_0)^{n_g-1}\right]^{k_{\rm max}}_{L^{-1}}/(n_g-1)
$
depends on wave numbers weakly, and
the scale dependence of $A$ 
can be safely neglected as long as CMB anisotropies (i.e. T and E) are concerned. 
In the following analysis, we simply replace $A$ with unity.
Then the reduced bispectrum in \eqref{eq:rb1}
can be rewritten as
\begin{equation}
b^{P_1P_2P_3}_{l_1l_2l_3} 
=\sum_{X_1X_2X_3}8\gamma_{X_1}\gamma_{X_2}\gamma_{X_3}\mathcal P_g(k_0)
[b^{X_1P_1,\,X_2P_2X_3P_3}_{l_1l_2l_3}+(\mbox{2 cyclic perms})].
\end{equation}
Here, $b^{X_1P_1,\,X_2P_2X_3P_3}_{l_1l_2l_3}$
is given as 
\begin{equation}
b^{X_1P_1,\,X_2P_2X_3P_3}_{l_1l_2l_3}
\equiv
\int r^2dr\,
\alpha^{X_1P_1}_{l_1}(r)
\beta^{X_2P_2}_{l_2}(r)
\beta^{X_3P_3}_{l_3}(r), 
\end{equation}
with $\alpha^{XP}_l(r)$ and $\beta^{XP}_l(r)$ being defined as
\begin{eqnarray}
\alpha^{XP}_l(r)&\equiv&
\frac{2}{\pi}\int k^2dk g^{XP}_l(k)j_l(kr), \label{eq:alphal} \\
\beta^{XP}_l(r)&\equiv&
\frac{2}{\pi}\int k^2dk P_g(k)g^{XP}_l(k)j_l(kr). 
\end{eqnarray}
For later convenience, we define the following set of non-linearity parameters
\begin{eqnarray}
f_{\rm NL}^{(1)}&=&8\gamma_1^3\mathcal P_g(k_0), \\
f_{\rm NL}^{(2)}&=&8\gamma_1^2\gamma_2\mathcal P_g(k_0), \\
f_{\rm NL}^{(3)}&=&8\gamma_1\gamma_2^2\mathcal P_g(k_0), \\
f_{\rm NL}^{(4)}&=&8\gamma_2^3\mathcal P_g(k_0).
\end{eqnarray}
Then the reduced bispectrum can be given as 
\begin{equation}
b^{P_1P_2P_3}_{l_1l_2l_3} 
=\sum_{j=1}^4 f_{\rm NL}^{(j)}
b^{P_1P_2P_3(j)}_{l_1l_2l_3}, 
\end{equation}
where $\{b^{P_1P_2P_3(j)}_{l_1l_2l_3}\}$
are template bispectra for the nonlinearity parameters $f_{\rm NL}^{(j)}$, which are defined as
\begin{eqnarray}
b^{P_1P_2P_3(1)}_{l_1l_2l_3}&=&b^{\mathcal R P_1, \mathcal R P_2\mathcal R P_3}_{l_1l_2l_3}+(\mbox{2 cyclic perms}), \\
b^{P_1P_2P_3(2)}_{l_1l_2l_3}&=&[b^{\mathcal S P_1, \mathcal R P_2\mathcal R P_3}_{l_1l_2l_3}
+b^{\mathcal R P_1, \mathcal S P_2\mathcal R P_3}_{l_1l_2l_3}+b^{\mathcal R P_1, \mathcal R P_2\mathcal S P_3}_{l_1l_2l_3}
]+(\mbox{2 cyclic perms}), \\
b^{P_1P_2P_3(3)}_{l_1l_2l_3}&=&[b^{\mathcal R P_1, \mathcal SP_2\mathcal S P_3}_{l_1l_2l_3}
+b^{\mathcal S P_1, \mathcal R P_2\mathcal S P_3}_{l_1l_2l_3}+b^{\mathcal S P_1, \mathcal S P_2\mathcal R P_3}_{l_1l_2l_3}
]+(\mbox{2 cyclic perms}), \\
b^{P_1P_2P_3(4)}_{l_1l_2l_3}&=&b^{\mathcal S P_1, \mathcal S P_2\mathcal S P_3}_{l_1l_2l_3}+(\mbox{2 cyclic perms}).
\end{eqnarray}

Now let us move on to the Fisher matrix analysis.
According to Refs.~\cite{Komatsu:2001rj,Babich:2004yc}, 
given template bispectra $b^{P_1P_2P_3(j)}_{l_1l_2l_3}$, 
the Fisher matrix for the nonlinearity parameters
$f^{(j)}_{\rm NL}$ can be approximately given as
\begin{eqnarray}
F_{jj'}&=&f_{\rm sky}
\sum_{l_1\le l_2\le l_3}
\frac{(2l_1+1)(2l_2+1)(2l_3+1)}{4\pi}
\begin{pmatrix}
l_1 & l_2 & l_3\\
0 & 0 & 0 
\end{pmatrix}^2\\
&\times&\sum_{P_1P_2P_3}
\sum_{P'_1P'_2P'_3}
b^{P_1P_2P_3(j)}_{l_1l_2l_3}
[\mathbf{Cov}_{l_1l_2l_3}^{-1}]^{P_1P_2P_3|P'_1P'_2P'_3}
b^{P'_1P'_2P'_3(j')}_{l_1l_2l_3},
\end{eqnarray}
where $f_{\rm sky}$ is the fraction of the sky covered by observations
and $\left[{\bf Cov}^{-1}_{l_1l_2l_3}\right]^{P_1P_2P_3|P'_1P'_2P'_3}$
is the inverse of covariance matrix.
The covariance matrix $\left[{\bf Cov}_{l_1l_2l_3}\right]^{P_1P_2P_3|P'_1P'_2P'_3}$
should be given in the limit of weak non-Gaussianity as
\begin{equation}
[{\bf Cov}_{l_1l_2l_3}]^{P_1P_2P_3|P'_1P'_2P'_3}
=\Delta_{l_1l_2l_3}\mathcal C^{P_1P'_1}_{l_1}
\mathcal C^{P_2P'_2}_{l_2}\mathcal C^{P_3P'_3}_{l_3},
\end{equation}
where $\mathcal C^{PP'}_l=C^{PP'}_l+N^{PP'}_l$
is the sum of signal ($C^{PP'}_l$) and noise ($N^{PP'}_l$) power spectra, and
$\Delta_{l_1l_2l_3}$ takes values 6, 2, 1 for the
cases that all $l$'s are the same, only two of them are the same
and otherwise, respectively.
For the noise power spectrum, we adopt the 
Knox's formula~\cite{Knox:1995dq},
\begin{equation}
N^{PP'}_l=\delta_{PP'}
\theta_\mathrm{FWHM}^2\sigma_P^2
\exp\left[l(l+1)
\frac{\theta_\mathrm{FWHM}^2}{8\ln2}
\right],
\end{equation}
where $\theta_\mathrm{FWHM}$ is
the full width at half maximum of the Gaussian
beam, and $\sigma_P$ is the root mean square 
of the instrumental noise par pixel.
For cases of multi-frequency observations, 
$N^{PP'}_l$ is given via quadrature sum the over frequency channels.
In Table~\ref{table:planck}, we summarized the survey parameters for Planck
that we adopt in what follows. In addition, we set the sky coverage $f_{\rm sky}$ to 0.7.

%%%
\begin{table}[htb]
  \begin{center}
  \begin{tabular}{l|c|c|c}
  \hline
  \hline
  bands [GHz] & $\theta_{\rm FWHM}$ [arcmin] & $\sigma_T$ [$\mu$K] & $\sigma_P$ [$\mu$K] \\
  \hline
      $30$ & $33.0$ & $2.0$ & $2.8$ \\ 
      $44$ & $24.0$ & $2.7$ & $3.9$ \\ 
      $70$ & $14.0$ & $4.7$ & $6.7$ \\
      $100$ & $10.0$ & $2.5$ & $4.0$ \\
      $143$ & $7.1$ & $2.2$ & $4.2$ \\
      $217$ & $5.0$ & $4.8$ & $9.8$ \\
      $353$ & $5.0$ & $14.7$ & $29.8$ \\
  \hline
  \hline 
\end{tabular}
  \caption{Survey parameters adopted in our analysis for Planck. 
  We here assume 1-year duration of observation.
  }
  \label{table:planck}
\end{center}
\end{table}
%%%

To translate the forecasted constraints on $f_{\rm NL}^{(j)}$ from the Fisher matrix
into those on $\gamma_1$, $\gamma_2$ and $\mathcal P_g(k_0)$,
we define an effective $\Delta \chi^2$ as follows
\begin{equation}
\Delta \chi^2
\equiv\sum_{jj'} f_{\rm NL}^{(j)}
F_{jj'}f_{\rm NL}^{(j')}.
\end{equation}
The expected allowed regions in the $\gamma_1$ and $\gamma_2$
plane are shown in Fig.~\ref{fig:chi2}.
Note that $\mathcal P_g(k_0)$ is degenerate with
$\gamma_1$ and $\gamma_2$, and we can constrain
only the combinations $\gamma_1\mathcal P_g(k_0)$ and
$\gamma_2\mathcal P_g(k_0)$. 
As can be read from the figure,  
when $\gamma_1=0$ and $\gamma_2=1$, $\mathcal P_g(k_0)=10^{-5}$ would be marginally 
allowed at 2 $\sigma$ level by Planck for $1.5\lesssim n_g\lesssim2$.

\begin{figure}
	\begin{center}
		\includegraphics[width=12cm]{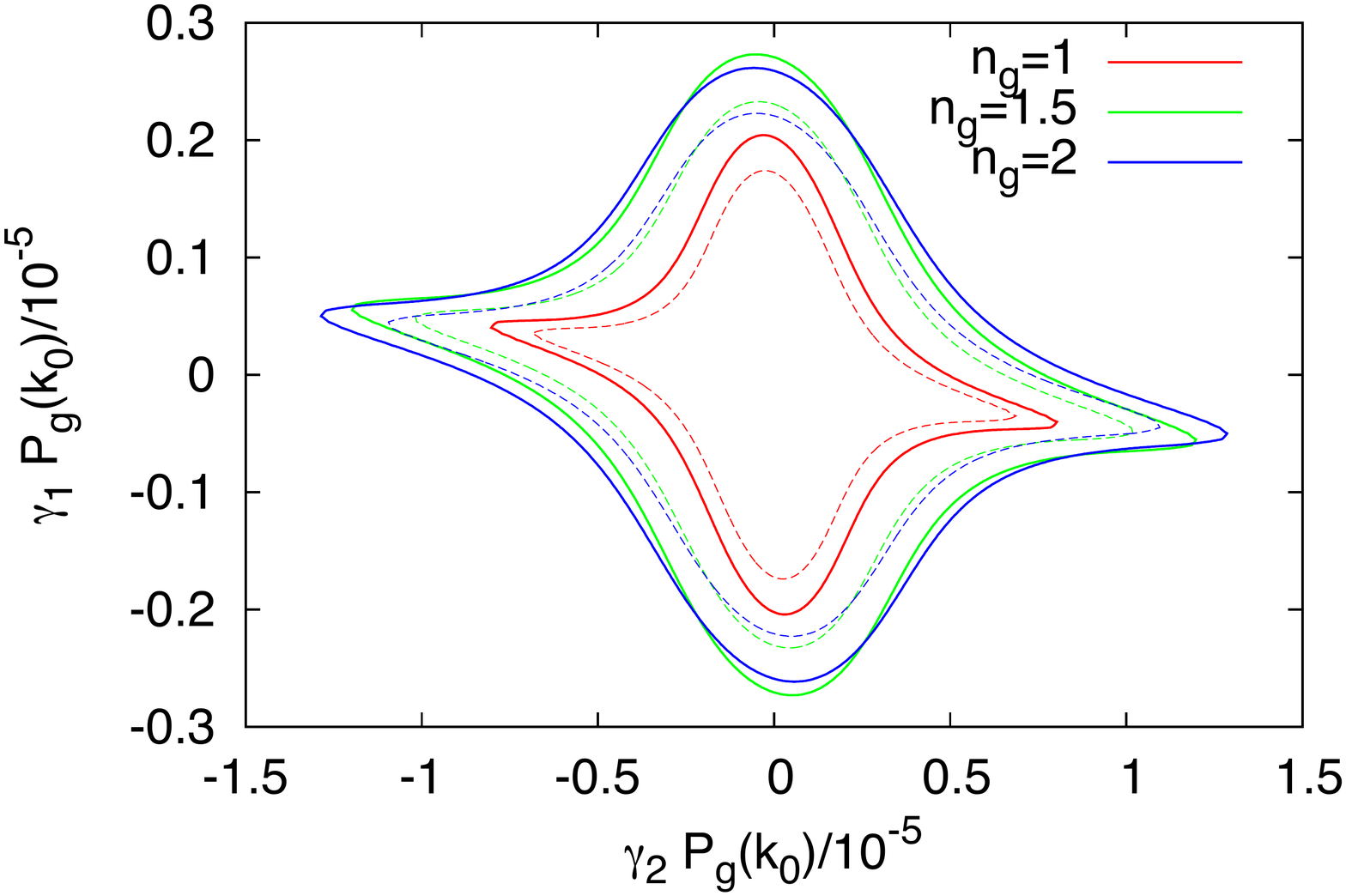}
		\caption{Forecasted constraints on $\gamma_1$ and $\gamma_2$
		in cases of  $n_g=1$ (red), 1.5 (green) and 2 (blue) are shown.
		Thick solid and thin dashed lines correspond to constraints at 1 and 2 $\sigma$ levels, respectively.
		}
	\label{fig:chi2}
	\end{center}
\end{figure}

In addition, we can also compute the expected constraint on $f_{\rm NL}^{{\rm loc,\nu}}$
from Planck by performing the Fisher matrix analysis in the similar manner.
Given the primordial bispectrum of \eqref{loc_nu}, 
the template bispectrum $\tilde b^{P_1P_2P_3, \nu}_{l_1l_2l_3}$ 
of the non-linearity parameter $f_{\rm NL}^{{\rm loc}, \nu}$
should be given as
\begin{equation}
\tilde b^{P_1P_2P_3, \nu}_{l_1l_2l_3}=-\frac65\int r^2dr\,
\tilde \alpha^{P_1}_{l_1}(r)
\tilde \beta^{P_2}_{l_2}(r)
\tilde \beta^{P_3}_{l_3}(r), 
\end{equation}
with $\tilde \alpha^{P}_l(r)$ 
and $\tilde \beta^{P}_l(r)$ being defined as
\begin{eqnarray}
\tilde \alpha^{P}_l(r)&\equiv&\alpha^{\mathcal SP}_l(r), \\
\tilde \beta^{P}_l(r)&\equiv&
\frac{2}{\pi}\int k^2dk P_{\mathcal S}(k)g^{\mathcal SP}_l(k)j_l(kr).
\end{eqnarray}
The reduced bispectrum is given by $
b^{P_1P_2P_3, \nu}_{l_1l_2l_3}=f_{\rm NL}^{{\rm loc}, \nu}
\tilde b^{P_1P_2P_3, \nu}_{l_1l_2l_3}$.
When $P_\mathcal S$ is scale-invariant, 
using the Planck survey parameters given in Table~\ref{table:planck}, 
we find the 1$\sigma$ error on
$f_{\rm NL}^{{\rm loc},\nu} \beta_{\rm iso}^2$ is expected to be 41.
Thus, if $\beta_{\rm iso}$ is assumed to be $10^{-1}$, $f_{\rm NL}^{{\rm loc}, \nu}$ can be 
as large as $10^4$ at around 2$\sigma$ level.

\end{document}